\documentstyle[aps,preprint,epsfig]{revtex}
\begin{document}
\topmargin=0.2in
          
\title{A Multi-Component Measurement of the Cosmic Ray\\ Composition 
Between ${\bf 10^{17}}$eV and ${\bf 10^{18}}$eV} 
\author{T.Abu-Zayyad$^1$, K.Belov$^1$, D.J.Bird$^{5}$, 
 J.Boyer$^{4}$, Z.Cao$^1$, M.Catanese$^3$, G.F.Chen$^1$,
 R.W.Clay$^{5}$, C.E.Covault$^2$, J.W.Cronin$^2$, H.Y.Dai$ ^1$,
 B.R.Dawson$^5$, J.W.Elbert$^1$, B.E.Fick$^2$, L.F.Fortson$^{2a}$,
 J.W.Fowler$^2$, K.G.Gibbs$^2$, M.A.K.Glasmacher$^7$,
 K.D.Green$^2$, Y.Ho$^{10}$, A.Huang$^1$ , C.C.Jui$^1$, M.J.Kidd$^6$,
 D.B.Kieda$^1$, B.C.Knapp$^4$, S.Ko$^1$, C.G.Larsen$^1$,
 W.Lee$^{10}$, E.C.Loh$^1$, E.J.Mannel$^4$, J.Matthews$^9$,
 J.N.Matthews$^1$ , B.J.Newport$^2$, D.F.Nitz$^8$,
  R.A.Ong$^2$, K.M.Simpson$^5$, J.D.Smith$^1$,
 D.Sinclair$^6$, P.Sokolsky$^1$, P.Sommers$^1$, C.Song$^4$, 
 J.K.K.Tang$^1$, S.B.Thomas$^1$, J.C.van der Velde$^7$,
 L.R.Wiencke$^1$, C.R.Wilkinson$^5$, S.Yoshida$^1$ and
 X.Z.Zhang$^4$ }

\address{$^1$ High Energy Astrophysics Institute, University of Utah, Salt Lake 
 City UT 8
 4112 USA\\$^2$ Enrico Fermi Institute, University of Chicago, Chicago IL 
 60637 USA\\
 $^3$ Smithsonian Astrophys. Obs., Cambridge MA 02138 USA\\
 $^4$ Nevis Laboratory, Columbia University, Irvington NY 10533 USA\\
 $^5$ University of Adelaide, Adelaide S.A. 5005 Australia \\
 $^6$ University of Illinois at Champaign-Urbana, Urbana IL 61801 USA\\
 $^7$ University of Michigan, Ann Arbor MI 48109 USA\\
 $^8$ Dept. of Physics, Michigan Technical University, 
 Houghton, MI 49931 USA\\
 $^9$ Dept. of Physics and Astronomy, Louisiana State 
 University, Baton Rouge LA 70803 and \\
 Dept. of Physics, Southern University, Baton Rouge LA 70801 USA\\
 $^{10}$ Dept. of Phys., Columbia University, New York NY 10027 USA\\
 $^a$ joint appt. with The Adler Planetarium and Astronomy Museum, 
 Astronomy Dept., Chicago IL 60605 USA\\
}
\date{\today}
\maketitle
\begin{abstract}
The average mass composition of cosmic rays with primary energies
between $10^{17}$eV and $10^{18}$eV has been studied using a
hybrid detector consisting of the High Resolution Fly's Eye
(HiRes) prototype and the MIA muon array. 
Measurements have been made of the change in the depth of
 shower maximum, $X_{max}$, and in the change in the muon density
at a fixed core location, $\rho_\mu(600m)$, as a function 
of energy. The composition has also been evaluated in terms of the
combination of $X_{max}$ and $\rho_\mu(600m)$.
The results show that the composition is changing from a heavy to
lighter mix as the energy increases.
\end{abstract}

%
%
The source of cosmic rays with particle energies above $10^{14}$ eV is 
still unknown. Models of origin, acceleration, and 
propagation must be evaluated in light of the observed energy spectrum
and chemical composition of the cosmic ray flux arriving at the earth. 
The cosmic ray energy spectrum
generally follows a simple power law over many decades of
energy. This might lead one to believe that cosmic
rays of all energies share the same source. However, there are
two detectable breaks in this otherwise smooth
spectrum. At an energy of about $10^{15}$ eV the spectrum softens.
At energies above $10^{18}$ eV it hardens
again. These two features, known as the ``knee'' and ``ankle'',
suggest that the source of cosmic rays or propagation effects 
might be changing in these regions.  
Observations of the mass composition as a
function of energy may provide a path to further understanding.

Several experiments have attempted to determine the mean cosmic
ray composition through the knee region of the spectrum.  While
the results are not in complete agreement, there is some
consensus for a composition becoming heavier at energies above
the knee \cite{BLANCA}, a result consistent with
charge-dependent acceleration theories or rigidity-dependent
escape models.
In the region above the knee, the Fly's Eye experiment has
reported a changing composition from a heavy mix around
$10^{17}$eV to a proton dominated flux around $10^{19}$eV
\cite{FEcomp}.  Muon data from the AGASA experiment show broad
agreement with this trend if the data are interpreted using the
same hadronic interaction model as in the Fly's Eye analysis
\cite{AGASA,DMS}.

%
%
In this experiment,  our
measurements are unique in that two normally independent
detection techniques are employed simultaneously in the
measurement of various aspects of extensive air shower(EAS). 
We use a hybrid detector consisting of the prototype High
Resolution Fly's Eye (HiRes) air fluorescence detector and the
Michigan Muon Array (MIA).  We have undertaken to independently measure
parameters reflecting the average cosmic ray nuclear
composition at energies above $10^{17}$ eV.
The detectors are located in the western desert of Utah, USA
at $112^{\circ}\,\mbox{W}\,\mbox{longitude}$ and $40^{\circ}
\,\mbox{N}\,\mbox{latitude}$.  The HiRes detector is situated
atop Little Granite Mountain at a vertical atmospheric depth of
$850$ g/cm$^2$. It overlooks the CASA-MIA arrays some 3.3
km to the northeast. The surface arrays are some 150 m
below the fluorescence detector at an atmospheric depth of 
870 g/cm$^2$.

The HiRes prototype has been described in detail elsewhere
\cite{hires}. It views the night sky with an
array of 14 optical reflecting telescopes. They image the
EAS as it progresses through the detection volume.  Nitrogen
fluorescence light (300--400\,nm) is emitted at an atmospheric
depth $X$ in proportion to the number of charged particles in the
EAS at that depth, $S(X)$. Part of this shower development profile
(at least 250 g/cm$^2$ long) can
be determined by measuring the light flux arriving at the detector.
Assuming $S(X)$ to be the Gaisser-Hillas~\cite{gh} shower developement 
function and correcting for Cherenkov light contamination and
atmospheric sccattering effects one can measure the
primary particle energy $E$, and the depth at which the
shower reaches maximum size, $X_{max}$\cite{Baltrusaitis}.

MIA~\cite{casamia}, consisting of over 2500 m$^2$ of 
active area distributed in 16 patches of 64 scintillation 
counters, measures EAS muon arrival times
with a precision of 4 ns and records all hits
occurring within 4 $\mu\mbox{s}$ of the system trigger.  MIA
records only the identification and firing time of the counters
participating in a given event. 
The average efficiency of MIA counters 
for detecting minimum ionizing particles was $93\%$ when they were 
buried, and the average threshold energy for vertical muons is 850 MeV.
MIA determines the muon density via the pattern of hit counters 
observed in the shower \cite{MIAanalysis}.  An estimate
of the muon density at 600\,m from the core, $\rho_\mu(600m)$, is
then determined by a fit.

%
%
It is expected that changes in the mean mass composition of the
cosmic ray flux as a function of $E$ will be manifested as
changes in the mean values of two measurable quantities 
${X}_{max}$ and ${\rho_\mu(600m)}$.
To indicate those changes,  a rate of change of
$X_{max}$ with $\log E$, called the elongation rate, $\alpha$,
has been introduced. Similarly for muons, we define a power law 
index for $\overline{\rho_\mu(600m)}$ as a function of $E$, 
called the ``$\mu$ content index'', $\beta$, in this study. Hence:
\begin{eqnarray}
\alpha = \frac{d\overline{X}_{max}}{d \log E}\ \ \ \ 
\rm{and} \ \ \ \beta = \frac{d\log \overline{\rho_\mu(600m)}}{d \log E}. 
\end{eqnarray}
Assuming that a shower initiated by a nucleus of
mass number $A$ and energy $E$ is a superposition of $A$
subshowers each with energy $E/A$, 
$\overline{X}_{max}\propto\alpha_0\log\left(E/A\right)$ 
and $\overline{\rho_\mu(600m)}\propto A\left(E/A\right)^{\beta_0}$
where $\alpha_0$ and $\beta_0$ are for a pure beam of primary nuclei 
of mass $A$. The values of $\alpha_0$ and $\beta_0$ 
are dependent of the  hadronic interaction model,
 but we find them largely 
independent of $A$ in our simulations described below.
Therefore, any deviation of our observed elongation rate, $\alpha$ and 
$\mu$ content index, $\beta$, from those for pure composition imply 
a changing  composition, i.e.
\begin{equation}
\frac{d \overline{\log A}}{d \log E}=-\frac{\alpha - 
\alpha_0}{\alpha_0}=\frac{\beta-\beta_0}{1-\beta_0} . 
\end{equation}
Since the superposition model is not fully realistic, 
a more reliable comparison between the data and 
predictions is based on detailed simulation of 
shower development described below. 

%
%
HiRes/MIA coincident data were collected on clear moonless nights
between Aug. 23, 1993 and Aug. 24, 1996. The total coincident
exposure time was 2878 hours corresponding to a duty cycle of
10.2\%. 4034 coincident events were observed.
 The shower trajectory for each event was obtained in an
iterative procedure using the information from both HiRes and MIA
\cite{Brian}. HiRes uses its spatial pixel patterns
to find the plane in space containing the the detector and the
shower axis (the shower-detector plane or SDP), and the time 
development to find the distance
of closest approach $R_p$. MIA helps to constrain the HiRes fits
using its muon arrival time distribution to provide the shower
direction within the SDP. The accuracy of the shower axis
determination depends on the
number of observed muons, the HiRes angular track length,
and the core distances from MIA and HiRes. Using its density 
measurement, MIA also helps to determine the shower core location 
in the SDP. 2491 events are 
reconstructed via this procedure.  Monte Carlo studies
\cite{OG1.3.05} show that over the full energy range studied,
the median shower direction error is $0.85^\circ$ with a median core
location error of 45\,m.

We have imposed cuts on the data to remove the poorly
reconstructed events and maintain good resolution.
The cuts are as
follows: triggered HiRes pixels should subtend an angle of at
least $20^\circ$ and view a depth range of at least
250\,g/cm$^2$; shower maximum should be bracketed by
measurements; the estimated error in shower maximum should be
less than 50\,g/cm$^2$; the reduced $\chi^2$ of the shower
profile fit should be less than 10; to reduce the influence of
direct Cerenkov light, all pixels should view the shower axis at
angles larger than $10^\circ$; and the reconstructed shower core
should be less than 2000\,m from the center of MIA.  These cuts
leave a sample of 891 events.  When determining the 
$\rho_\mu(600m)$ from muon data we also require that: the
shower core lie between 300\,m and 1000\,m from MIA; and
the number of hit MIA counters should be less than 700 to avoid 
the saturation of the $\mu$ counters.  This
more restricted sample contains 573 events.

\begin{figure}[t]
\epsfig{file=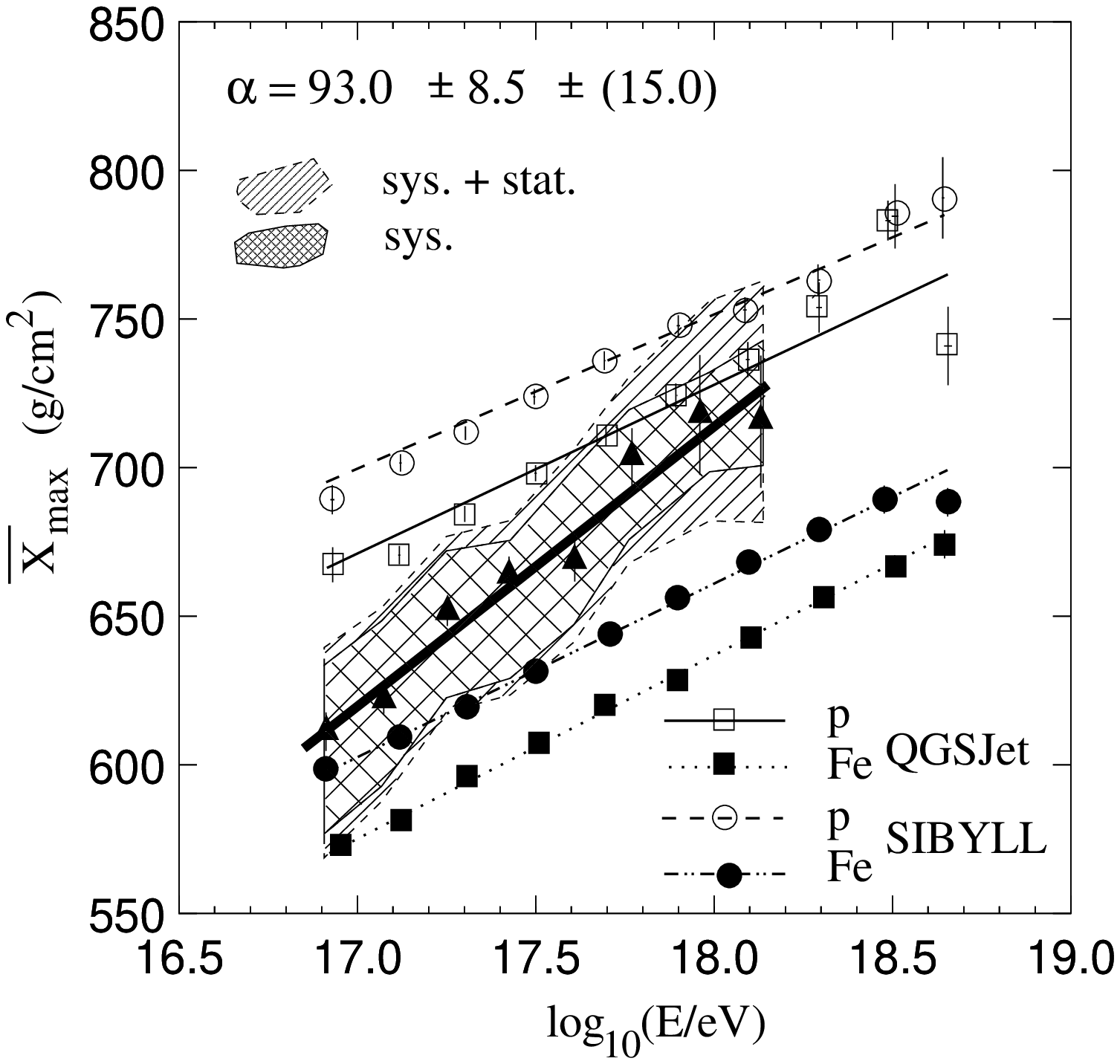}
\caption{Average $X_{max}$ increasing with energy. Shaded areas and the thick 
line within the area represent HiRes data and the best fit of the data respectively. 
The closed triangles represent the data set corresponding to the central values of 
the parameters in the reconstruction. 
The circles, squres and lines refer to the simulation results. 
See text for details.
\label{fig1}}
\end{figure}

\begin{figure}[t]
\epsfig{file=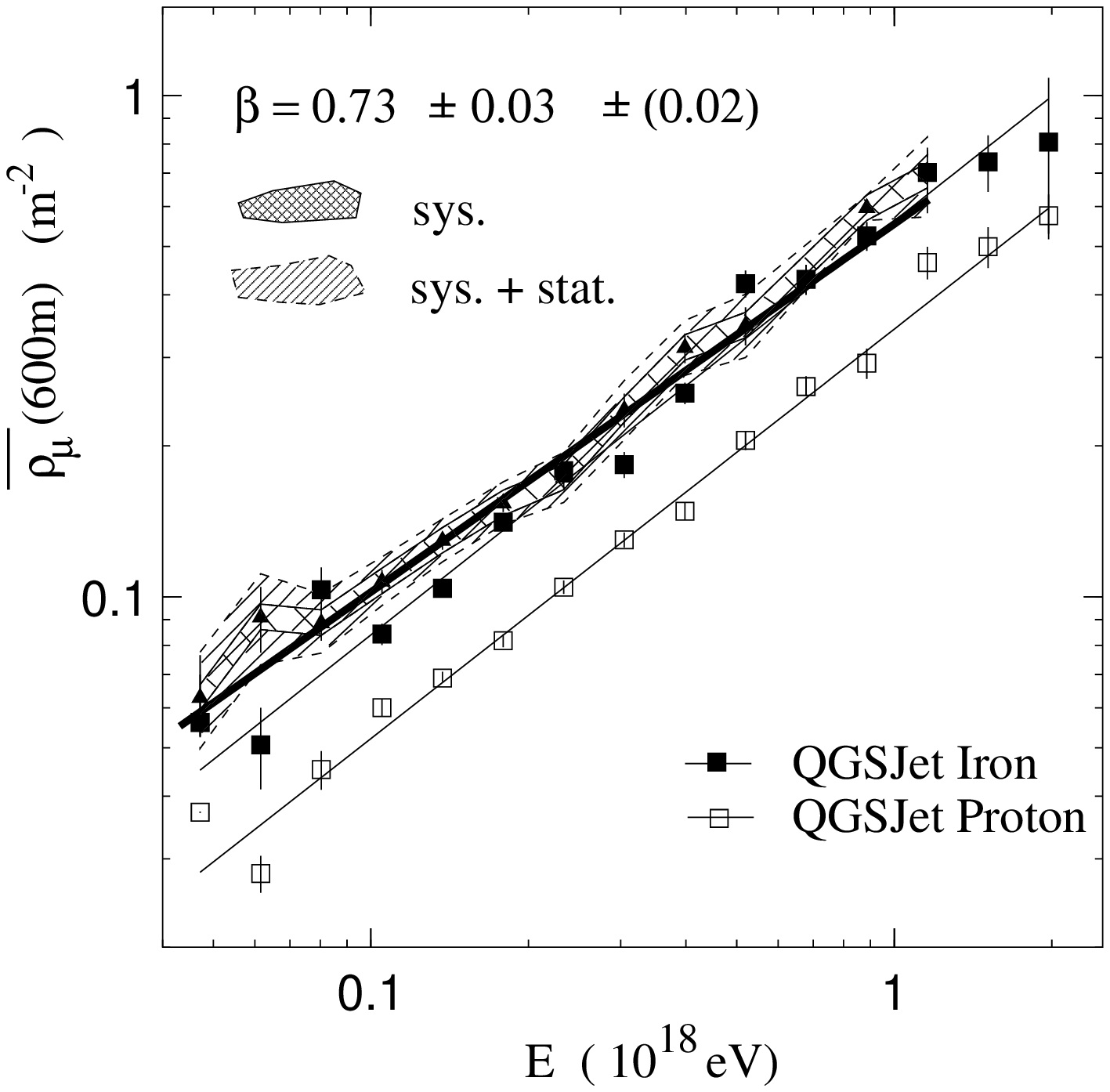}
\caption{Average Muon density at 600 m from the shower core. Same as FIG 1. 
\label{fig2}}
\end{figure}

%
%
The HiRes data are shown in FIG.~\ref{fig1}. We show bands
to represent the statistical and systematic uncertainties in the
depth of maximum. 
The measured elongation rate is
$93.0\pm 8.5\pm (10.5)$\,g/cm$^2$/decade 
over the energy range from $10^{17}$ to $10^{18.1}$ eV. 
The MIA data are shown in FIG.~\ref{fig2}. The
data show a $\mu$ content index of $0.73\pm 0.03\pm (0.02)$/decade
in the same energy range as for the elongation rate. Numbers in the 
brackets provide the systematic error based on the analysis described 
below. 
 
The systematic errors on the  muon local density $\rho_\mu(600m)$ 
result from the 
uncertainties in the absolute efficiencies of the MIA counters
over time. The average efficiency is 80.7\% 
with an RMS of 4.7\% over the 16 patches during the time 
the data were taken.
This is the only significant systematic uncertainty associated 
with $\rho_\mu(600m)$.
For $X_{max}$, we have considered systematic errors in the
atmospheric transmission of light and in the production of
Cerenkov light.  These are related since atmospherically
scattered Cerenkov light can masquerade as fluorescence light if
not accounted for properly.  For atmospheric scattering, there
was uncertainty in the aerosol content in the air from night
to night.  The uncertainty, equivalent to one standard deviation 
about the mean, is expressed as a range of possible
horizontal extinction lengths for aerosol scattering at 350\,nm
(taken as 11\,km to 17\,km based on measurements using Xenon flashers)
\cite{flasher}
and a range of scale heights for the vertical
distribution of aerosol density above the mixing layer (taken as
0.6\,km to 1.8\,km).  For Cerenkov light production, we have
systematically varied the angular scale for the Cerenkov emission
around the shower axis.  The emission angle distribution is
related to the angular distribution of the shower particles and
the intrinsic Cerenkov emission angle.  Both vary with depth in
the atmosphere, and at ground level we take the distribution as
an exponential function of the angle from the shower axis, with a
scale of $4.0\pm0.3^\circ$. Those uncertainties are shown by the 
shaded areas in FIG 1 \& 2.

Another systematic error, that in the determination of energy, is 
considered to be independent of energy and will therefore have no effect
on the values of the measured elongation rate and $\mu$ content index. It will
however effect the normalization of the mean depth of maximum or muon size
at a particular energy.  We estimate that the systematic error in energy 
is no larger than 25\%, made up of an uncertainty in the nitrogen 
fluorescence efficiency of no more than 20\% and a calibration systematic 
of less than 5\%.

%
%
Also shown in FIG.~\ref{fig1}\ and \ref{fig2}\ are Monte Carlo 
simulation results. 
These full shower simulations have been performed using the
CORSIKA package \cite{CORSIKA}, employing QGSJET\cite{QGSJET}
 and SIBYLL\cite{SIBYLL} hadronic
interaction models.  We have generated showers
of fixed energies and fixed zenith angles in order to parameterize
the shape of the shower development profile and the muon content
at any energy between $3\times10^{16}$eV and $5\times10^{18}$eV and
at any zenith angle out to $80^\circ$. We then pass those showers
through a realistic simulation of the detector and generate data
to be analyzed by the reconstruction software used for real data.
A $E^{-3}$ differential spectrum is assumed.
The minimum energy is well below the HiRes/MIA threshold to allow
for a study of any threshold effects.  Distributions of energy,
impact parameter and zenith angle are well predicted by the
simulation \cite{OG1.3.05}.

After reconstruction, and after applying the same quality cuts as
we apply to the real data, we find that a pure proton flux and
the QGSJET model gives an elongation rate of
$\alpha_0=58.5\pm1.3$\,g/cm$^2$/decade and a $\mu$ content index of 
$\beta_0=0.83\pm 0.01$/decade over the range from $10^{17}$ to
$10^{18}$eV. For a pure iron composition and the QGSJET model we
find corresponding values of $\alpha_0=60.9\pm1.1$\,g/cm$^2$/decade 
and the same $\beta_0=0.83\pm0.01$\,g/cm$^2$/decade as for protons.  
Results from SIBYLL show similar
elongation rates, but have the $X_{max}$ approximately
25\,g/cm$^2$ deeper than QGSJET.  
SIBYLL also predicts significantly fewer
muons at 600\,m for both proton and iron showers.  The effect of any
triggering and reconstruction biases is very small for $X_{max}$, 
as can be
seen in FIG.\ref{fig1} by comparing these reconstructed data (dots) with the 
``input'' (lines) directly from CORSIKA. 
The application of well chosen
cuts has resulted in a bias-free measurement of the elongation rate.
         However, for muon 
density measurement, reconstruction effects change the index by 8\%. The
``input'' for the simulation is 
0.88 (true QGSJET prediction) for both proton and 
iron showers. 
We suspect that the presence of an asymmetry in core distance error 
can result in a small overestimate of the muon density. 
It is  possible that this effect changes with shower energy.

Data and simulations are clearly 
inconsistent with each other in both the elongation rate and the $\mu$
content index. It leads support to the hypothesis that
the cosmic ray composition is changing towards a lighter mix of nuclei
from $10^{17}$
to $10^{18}$eV. HiRes and MIA reach the same 
conclusion by 
using different experimental techniques and measuring different physics 
variables. Substituting the  measured and simulated values of 
$\alpha$ and $\beta$ in (2) shows that the result 
from HiRes and MIA are consistent, with an implied change in
 $\Delta\overline{\log A}$  
of about -0.58 over one decade of energy. 

There is a problem with the absolute density of muons at 600\ m, 
with respect to the model predictions.  As seen in
FIG.~\ref{fig2}, the data show values of $\rho_\mu$ at the lower
energies which are larger than pure iron showers.  The comparison
with SIBYLL predictions (not shown) is even worse.  The reason
for this discrepancy is not clear.  We have searched for a
possible experimental reason for a systematic overestimation of
the muon size, and have not found an affect of sufficient magnitude.  
A change in the energy scale by the maximum estimated systematic error
of 25\% world lessen the problem but not completely remove it.
A shift of 40\% is required, which is well beyond the systematic error.
such a shift would also produce large discrepancies in the R$_p$ 
distribution between data and Monte Carlo.
We conclude that the model is deficient in muon production.  Evidence
for this has also been found at lower EAS energies \cite{KASCADE}. The 
local density is consistent with AGASA experiment, e.g.
$\rho_\mu(600m)=0.24\pm 0.02\pm (0.02)$/m$^2$ for this work and 
$\sim 0.25$/m$^2$ for AGASA~\cite{AGASA} at $3\times 10^{17}$eV. 
 Since models with very different predictions of 
$\rho_\mu(600m)$ have very similar $\mu$ content indices, we believe 
that the $\mu$ content indices are more reliable 
 than the muon normalization.

\begin{figure}[b]
\epsfig{file=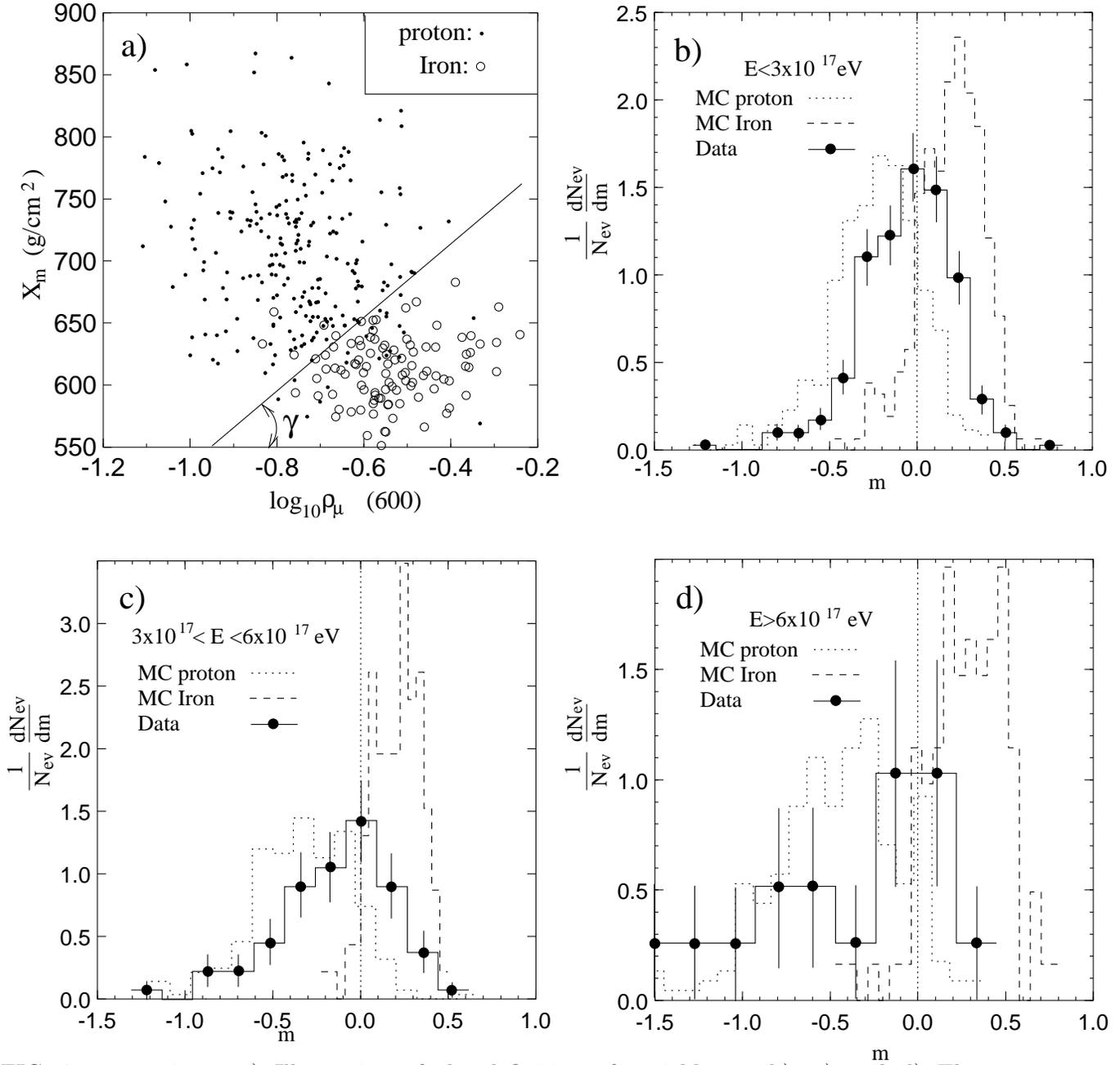, width=18cm}
\caption{$m$ testing. a) Illustration of the definition of variable $m$; b), c) and  d) 
The $m$-distributions and the comparison with the simulations in three energy regions.
\label{fig3}}
\end{figure}

%
%
We can take full advantage of the hybrid nature of our experiment
by constructing a parameter which is a combination of the $X_{max}$
and $\rho_\mu(600m)$ measurements. Based on the simulation with  
fully considered shower development and detector responses, 
a new parameter can be determined by
first finding a line in $X_{max}$-$\log\rho_\mu(600m)$ space
which efficiently separates simulated proton and iron showers
(FIG.~\ref{fig3}(a)).  The dimension perpendicular to this line, 
called $m$, can be defined as
$m=\cos\gamma\log\rho_\mu(600m)$-$\sin\gamma X_{max}+m_0$, 
where $\gamma$ refers 
to the angle between the line and the log$\rho_\mu(600m)$ axis and $m_0$ 
is an arbitrary parameter to make $m>0$ for iron and $m<0$ for proton. 
The value of $m$ for a given shower indicates the degree of 
similarity between the data and either a pure iron or a pure proton 
event. FIG.~\ref{fig3}(a)
shows that the separation between proton and iron showers is
larger in terms of $m$ than it is for either $X_{max}$ or
$\rho_\mu(600m)$, thus the fluctuations from event to event may be
useful as an indication of the composition.

We plot the distribution of $m$ for three different energy ranges
in FIG.~\ref{fig3}. We compare this for the data (solid lines) with 
the estimates for simulated fluxes of pure protons (dashed lines) 
and pure iron (dotted lines) under the QGSJET assumption with full
detector simulation and reconstruction.  
All the distributions are normalized. It is obvious 
that neither pure proton nor 
pure iron can account for the data for energies under $3\times10^{17}$ eV. 
The data are
highly peaked at $m=0$, implying a mixed composition
 around $10^{17}$ eV. The proportion of proton like events 
increases with energy.
Note that due to the discrepancy regarding the muon normalization 
between data and simulation, we have shifted the log$\rho_\mu(600m)$ values
 by adding 0.17 to the predictions. This makes the muon results consistent 
with the $X_{max}$ results in terms of the normalization.

Thus FIG.~\ref{fig3} shows a change in the mean value of $m$ 
indicating a lightening of the mean cosmic ray mass.  In
the three energy ranges indicated, $m$ has values of $-0.061\pm0.015$,
$-0.148\pm0.036$ and $-0.42\pm0.25$. For comparison, $m$
for pure proton changes from -0.23$\pm$0.01 to -0.46$\pm$0.02
 while $m$ for pure iron is relatively stable around 0.2.

%
%
We conclude that the HiRes-MIA hybrid experiment confirms
the Fly's Eye experiment result that the primary composition 
changes towards a lighter mix of nuclei from $10^{17}$ to $10^{18}$eV.
This confirmation is nontrivial not only because of its unique 
combination of the simultaneous observation of shower longitudinal 
development and muon density on the ground, but also because of
its significantly improved $X_{max}$ and energy resolution. 
While the conclusion regarding the primary composition 
depends on the interaction model used, this study shows
that the elongation rate is relatively stable with respect 
to choice of models. No modern interaction model has produced 
an $\alpha_0$ much larger than 60 g/cm$^2$/decade.

A change from a heavy to a light composition in this energy region 
may indicate the increasing abundance of extra-galactic cosmic rays. 
Indeed, the Flys's Eye experiment \cite{FEcomp} reports a change in the 
spectral index near $5\times10^{18}$eV
which can be interpreted in this light. 
A number of new experiments, such as HiRes, the Pierre Auger Project 
and the Telescope Array, 
could address this issue. However, the source of the 
lower energy heavy composition remains a mystery. In that regard we note that 
there is a need to explore the energy region between  $10^{16}$ to $10^{17}$ eV
in order to 
connect our results with the measurements performed below $10^{16}$ eV. A  
measurement of the composition in this region may be crucial for the 
understanding of the sources of cosmic rays above the ``knee''.



\end{document}